\documentclass[preprint,prd,nofootinbib,onecolumn,showpacs]{revtex4}
\usepackage{dcolumn}
\usepackage{lipsum}
\usepackage{graphicx,epsfig}
\usepackage{psfrag}
\usepackage{bm}
\usepackage{epstopdf}
\usepackage{latexsym}
\usepackage{multirow}
\usepackage{amsmath,amssymb}
\usepackage{enumerate}
\usepackage[FIGTOPCAP]{subfigure}

\def\be {\begin{equation}}
\def\ee {\end{equation}}
\def\ba {\begin{eqnarray}}
\def\ea {\end{eqnarray}}
\def\nn {\nonumber}
\def\bc {\begin{center}}
\def\ec {\end{center}}
\def\b  {\beta}
\def\c  {\gamma}

\newcommand{\bdm}{\begin{displaymath}}
\newcommand{\edm}{\end{displaymath}}

\def\O  {\Omega}
\def\p  {\pi}

\def\r  {\rho}

\def\la {\label}
\def\le {\left}
\def\ri {\right}
\def\pa {\partial}
\def\f {\frac}
\def\sq {\sqrt}

\def\bi {\begin{itemize}}
\def\ei {\end{itemize}}

\def\bc {\begin{center}}
\def\ec {\end{center}}
\def\vph {\varphi}

\begin{document}

\title{Entanglement entropy due to near horizon degrees of freedom}

\author{Suman Ghosh} \email[email: ]{suman.ghosh@bose.res.in}
\affiliation{Department of Theoretical Sciences, S N Bose National Centre for Basic Sciences, Kolkata - 700098, India}

\begin{abstract}
Assuming that the dominant contribution, to the entropy due to entanglement across a spherical hypersurface, comes from the near horizon degrees of freedom, we analytically derive the entropy of a free massless scalar field in Minkowski spacetime across a spherical entangling surface. 
The resulting entanglement entropy is found to be proportional to the entangling surface as expected. A logarithmic subleading term with positive coefficient is also found through numerical computation. We have extended the analysis to higher dimensions as well.

\end{abstract}

\maketitle

\section{Introduction}



There are many approaches to quantum theory of gravity where Bekenstein-Hawking entropy ($S_{BH}$) of black holes \cite{EE-BH} can be derived \cite{York:1983zb,thooft-brickwall-1985,Frolov:1993ym,Strominger:1996sh, lqg-SBH,Carlip:2002be,Carlip:2000nv,Kaul:2000kf,Sen:2007qy}. In many cases they lead to power-law and logarithmic corrections.
Ever since Bombelli et al \cite{1986-Bombelli.etal-PRD} and Srednicki \cite{1993-Srednicki-PRL} showed that entanglement entropy (EE) of a massless free scalar field, in it's ground state in flat spacetime, is proportional to the area of the horizon, EE is considered as one of the most promising candidates as a source of $S_{BH}$ or a quantum correction to the same. EE is defined as the entropy due to entanglement between degrees of freedom (DoF) on the two sides of an entangling surface (or the so-called horizon).
One can argue that the computation of EE in \cite{1986-Bombelli.etal-PRD,1993-Srednicki-PRL} did not involve black hole geometry as such. However, as shown in \cite{Das:2007mj} certain modes of gravitational perturbations in black-hole space-times behave as minimally coupled scalar fields. Further, the Hamiltonian for a scalar field in Schwarzschild background can be shown (using general linear transformations in Lamaitre coordinates) to be equivalent to that in a flat spacetime \cite{Das:2007sj}. EE in fact takes into account the most important physical effect of an event horizon, that is to block information to an outside observer. Using this so-called real time approach or non-geometric approach, it has also been shown that EE in presence of excited and mixed states \cite{Das:2007mj} 
 lead to power-law corrections \cite{Shankaranarayanan:2011sd}.
In \cite{Theisen-2009}, a logarithmic correction is found by numerically fitting the non-geometric EE and the resulting coefficient was in agreement with that predicted by geometric approaches \cite{Solodukhin}.

In recent years EE is found to be playing crucial roles in understanding many quantum phenomena and their applications \cite{2009-Horodecki.etal-RMP,2010-Eisert.etal-RMP,2008-Das.etal-PRD}. Deriving EE, in non-geometric approach, analytically in 3D (3 space and 1 time dimension) field theory is difficult and exact results has only been found numerically. 
However, EE has been derived analytically (for Rindler horizons) using path integral methods \cite{Kabat-Strassler-1994,Kabat-1995} and in the context of 2D and 4D conformal field theories (as the so-called geometric entropy) using the replica method \cite{Holzhey-etal-1994, Callan-1994, Cardy-Calabrese-2004,Vidal-etal-2013,Latorre-etal-2004}. This method is also applied to compute EE for horizons with conical singularities \cite{Solodukhin,Casini-2006,Casini-2010}-- a logarithmic correction term is found in even spacetime dimensions. 
The holographic definition of EE \cite{Ryu-Takayanagi-prl-2006} is an exciting proposal and further attempts are being made to understand its implications \cite{Ryu:2006ef,Casinietal-2011,LM-2013}. From an information theoretic perspective, Plenio et al \cite{Plenio:2004he} have found the bounds on EE analytically in case of a 3D Cartesian lattice and planar entangling surfaces.

It is widely accepted that EE obeys the so-called area law (in case of $n-sphere$ in flat spacetime). From dimensional arguments \cite{Casini-2006}, one can write down the subleading terms too. 
The computational algorithm in the non-geometric approach as presented in \cite{1986-Bombelli.etal-PRD,1993-Srednicki-PRL}, is straightforward and unambiguous though, is impossible to be carried out completely analytically. Remarkably, the output of all the complicated numerical evaluation is a simple area law. This observation indicates that it might be possible to derive the dominant term which is proportional to the area {\em analytically} using some reasonable approximations. 
It is shown in \cite{Das:2007sj} that the DoF near the entangling surface contributes the most to the total entropy. Thus to be able to find the leading term, it is appropriate to consider entanglement among the near horizon DoF only while neglecting the rest. 
With this reasonable approximation, we show that, the resulting EE, derived analytically, follows an area law.
Let us define, $EE_{tot}$ to be the EE due to all the entangled DoF inside and outside the horizon and $EE_{surf}$ is defined as the EE due to correlation among the near horizon DoF that are residing just across the horizon. Arguably, the DoF responsible for entropy of a black hole also resides near the black hole horizon \cite{Wald:1999vt}. In this sense, $EE_{surf}$  is the dominant quantity in the context of $S_{BH}$ and here we compute the same analytically. 

This article is organised as follows. In Section II, we briefly review the standard algorithm \cite{1986-Bombelli.etal-PRD,1993-Srednicki-PRL} to compute EE for a real free massless scalar field propagating in $(3+1)$-dimensional flat space-time (for spherical horizon). In Section III, we analytically derive $EE_{surf}$ which is proportional to area. 
We also cross-check our result through numerical computations that further reveals a positive logarithmic correction. We compare these results regarding $EE_{surf}$ with $EE_{tot}$ in the lights of \cite{Theisen-2009} to determine the role of the DoF away from the horizon.
We generalise our calculations for spherical entangling surfaces in any dimension.
Then the ratio of $EE_{surf}$ and $EE_{tot}$, computed numerically, is shown to be tending to unity with increasing dimension of space. 
Finally, we summarise with a discussion on the implications of our results and related open issues in Section V.


\section{Brief review of the computational algorithm}

The Hamiltonian is given by
\ba
H &=& \f{1}{2} \int d^3x \le[ \p^2(x) + |\vec\nabla\varphi(\vec x) |^2 \ri].  \la{eq:Ham}
\ea
%

Decomposing the field and its conjugate momentum in partial waves
{\small
\ba
\varphi(\vec r)  = \sum_{\ell m} \f{\varphi_{\ell m}(r)}{r}~Y_{\ell m} (\theta,\phi)~,~ 
~\pi(\vec r)  = \sum_{\ell m} \f{\pi_{\ell m}(r)}{r}~Y_{\ell m} (\theta,\phi) \nn
\ea
}
where $Y_{\ell m}$'s are real spherical harmonics. The operators defined above are Hermitian and obey the appropriate commutation relations.
Integrating over $\theta$ and $\phi$ directions yield:
%
\be
H =\sum_{\ell m} \f{1}{2} \int_0^\infty dr
\le[\p_{\ell m}^2(r) + r^2 \left( \f{\pa}{\pa r} \le( \f{\varphi_{\ell m} (r)}{r}\ri) \ri)^2 
 + \f{\ell(\ell+1)}{r^2} \varphi_{\ell m}^2(r) \ri] 
\la{eq:Ham2}
\ee
To regularize, one discretizes the Hamiltonian (\ref{eq:Ham2}) along the radial direction with lattice spacing $a$, such that $r \rightarrow r_j = ja;~r_{j+1}-r_j=a$. This implies that contributions of the modes with linear momentum above $a^{-1}$ are exponentially suppressed. The lattice is terminated at a large but finite $N$ (we have chosen $N=100$ for numerical computations). An intermediate point $n$ is chosen, such that $n+\f{1}{2}$ represents a point on the (imaginary) spherically symmetric entangling surface or the {\it horizon} with radius ${\cal R}$ ($= a(n+\f{1}{2})$), that separates the lattice points between the {\it inside} and {\it outside}.  

After discretization, one can map Eq. (\ref{eq:Ham2}) with the Hamiltonian of N coupled harmonic oscillators written as
{\small
\be
H = \sum_{j} H_{j} = \frac{1}{2a} \sum_{i,j}^{N} \delta_{ij} \pi_{j}^2 + \vph_{j} \, K_{ij} \, \vph_j \,  \equiv \frac{1}{2} \sum_{i,j}^{N} \delta_{ij} p_{j}^2 + r_i \, K_{ij} \, r_j .
\label{eq:discretizeHam}
\ee } 
Here $\vph_j$'s, $\pi_j$'s and $K_{ij}$'s are dimensionless and the interactions are contained in the off-diagonal elements of the matrix $K_{ij}$ whose non-zero elements are given by,
\ba
K_{jj} = 2 + \f{1}{2j^2} + \f{\ell(\ell+1)}{j^2}; ~ K_{j,j+1} =K_{j+1,j} = -\f{(j+\f{1}{2})^2}{j(j+1)}.
 \la{eq:Kij}
\ea
Note that, Eq. (\ref{eq:Kij}) always satisfies the condition, for positivity of the eigenvalues \cite{Plenio:2004he}.
%

A brief description on how to calculate entropy from the above Hamiltonian is the following. The reduced density matrix (for ground state), tracing over the first $n$ of $N$ oscillators, 
is given by:
%
\be
\rho_{\rm red} = 
\int \prod_{j=1}^n~dr_j~\varphi_0 (r_1,\dots,r_n;r_{n+1},\dots,r_{N})~ 
\times  \varphi_0 (r_1,\dots,r_n;r'_{n+1},\dots,r_{N}') 
\la{den2} 
\ee
where $r$ and $r'$ represent radial distances outside the horizon from the center.
The resulting $\rho_{_{\rm red}}$ is a mixed state of a bipartite system. Entanglement is computed as the von Neumann entropy associated with the reduced density matrix $\rho_{_{\rm red}}$ \cite{2009-Horodecki.etal-RMP,2010-Eisert.etal-RMP}:
\be
S = - \mbox{Tr} \, \left( \rho_{_{\rm red}} \ln \rho_{_{\rm red}} \right)
\ee
The ground state is
\be
\varphi_0(r_1,\dots ,r_N) = 
\le(\f{\det \Omega}{\pi^N}\ri)^{\f{1}{4}} \exp\le[ -\f{1}{2} r^T\Omega r \ri]
\la{gs2}
\ee
where $\O$ is the square root of $K$ i.e. if $K =U^T K_D U$ where $K_D$ is diagonal and $U$ is a orthogonal matrix then $\O = U^T K_D^{1/2}U$. Corresponding density matrix (\ref{den2}) can be written as
%
\be
\r_{\rm red}  \sim \exp\le[ -(r^T \gamma r + r'^T\gamma r')/2 + r^T\beta r' \ri] 
\la{gsden}
\ee
where:
\be 
\O = \le( \begin{array}{ll} 
A& B  \\ 
B^T & C \\
\end{array} \ri)~,~\beta=\f{1}{2}B^TA^{-1}B~,~\gamma=C-\beta~  \la{eq:beta-gamma}.
\ee
%
The Gaussian nature of the above density matrix lends itself to a series of diagonalizations
\be
V\c V^T = \c_D, \b' \equiv \c_D^{-\f{1}{2}} V \b V^T \c_D^{-\f{1}{2}}, W  \b' W^T = \b'_D, v_j \in v \equiv W^T (V \c V^T)^{\f{1}{2}} VT \la{eq:diag} 
\ee
such that it reduces to a product of $(N-n)$, $2$-oscillator density matrices, in each of which one oscillator is traced over:
\be
\rho_{_{\rm red}}  \sim
\prod_{j=1}^{N-n} 
\exp\le[-\f{v_j^2+v_j'^2}{2} + \beta'_j v_j v_j' \ri] ~.
\ee
The corresponding entropy is given by:
\be
S_{\ell} = \sum_{j=1}^{N-n} 
\le( - \ln[1-\xi_j] - \f{\xi_j}{1-\xi_j}\ln\xi_j  \ri) \la{eq:S_l}
\ee
where
\be
\xi_j = \f{{\b}_j'}{1+ \sqrt{1-{\beta}_j'^2}}. \la{eq:xi_l}
\ee
Thus, for the full Hamiltonian $H=\sum_{\ell m}H_{\ell m}$, the entropy is:
\be
EE = \sum_{\ell=0}^{\infty} (2 \ell+ 1) S_{\ell}, \la{eq:S_tot}
\ee
where the degeneracy factor $(2\ell+1)$ follows from spherical symmetry of the Hamiltonian. 
%
Note that the above sum will not converge in dimensions larger than four.

\section{Analytic computation of $EE_{surf}$}

 
As we mentioned earlier, $EE_{surf}$ can be regarded as the  most dominant contribution to the $EE_{tot}$. 
Here we compute $EE_{surf}$ analytically using the same algorithm presented in the previous section. To take into account only the near horizon DoF, one needs to make all the off-diagonal terms in $K_{ij}$, except those that correspond to the interaction between $n$-th and $(n+1)$-th lattice sites, vanish.  Schematically, $K_{ij}$ ($\equiv K^i_j$) in Eq. (\ref{eq:Kij}) simplifies as follows:
{\small
\ba
K^i_j\longrightarrow
\le( \begin{array}{llllll} 
{\times} & {} & {} & {} & {} & {} \\
{} & {K^{n-1}_{n-1}} & {} & {} & {} & {} \\
{}  & {} & {K^n_n} & {K^n_{n+1}} & {} & {} \\
{} & {}  & {K^{n+1}_{n}} & {K^{n+1}_{n+1}} & {} & {} \\
{} & {} & {} & {} & {K^{n+2}_{n+2}} & {}  \\
{} & {} & {} & {} & {} & {\times}  \\
\la{mat2} \\ 
\end{array} \ri) \label{eq:Kij_simp}
\ea
}

where the `$\times$' represents rest of the diagonal elements. Note that, for $EE_{tot}$, all the elements in the first off-diagonals in $K_{ij}$ are non-zero.
Let us only consider `large' entangling surfaces so that $n\gg 1$. Further, as is shown later, the $\ell \sim n$ modes contributes dominantly to the entropy whereas contribution of lower modes are negligible. Thus for $\ell \gg 1$, we have
\be
K^n_n \sim K^{n+1}_{n+1} \sim 2 + \f{\ell^2}{n^2};\,\, K^n_{n+1} = K^{n+1}_n \sim -1 \la{eq:Knn}
\ee
This leads to
{\small
\ba
\O^i_j  = 
\le( \begin{array}{llllll} 
{\sqrt{\times}} & {} & {} & {} & {} & {} \\
{} & {\sqrt{K^{n-1}_{n-1}}} & {} & {} & {} & {} \\
{}  & {} & {K_+} & {K_-} & {} & {} \\
{} & {}  & {K_-} & {K_+} & {} & {} \\
{} & {} & {} & {} & {\sqrt{K^{n+2}_{n+2}}} & {}  \\
{} & {} & {} & {} & {} & {\sqrt{\times}}  \\
\la{matOmega} \\ 
\end{array} \ri) 
\ea
}
where
\be
K_{\pm} = \f{\sqrt{K^n_n-1} \pm \sqrt{K^n_n+1}}{2}
\la{eq:K}
\ee

Using Eq. (\ref{eq:beta-gamma}) we get the only non-zero component of $\b$ (as $B_{jn} = 0 ~ \forall ~j \neq 1$) given by,
\be
\b_{11} = \f{1}{2}B_{1n} A_{nn}^{-1} B_{n1} = \f{K^n_n - \sq{(K^n_n)^2 - 1}}{2(\sq{K^n_n-1} + \sq{K^n_n-1})}
\ee
Eq. (\ref{eq:diag}) further implies that there is only one non-zero eigenvalue of $\beta'$, given by
\ba
\b' \equiv \f{\b_{11}}{\gamma_{11}} &=& \f{\b_{11}}{C_{11}-\b_{11}} =  \le[4K_{nn}(K_{nn} + \sqrt{K_{nn}^2 - 1}) - 3\ri]^{-1}. \\
&=& \le[4 \le(2+ \f{\ell^2}{n^2}\ri) \le(2 + \f{\ell^2}{n^2} + \sqrt{\f{\ell^4}{n^4} + 2\f{\ell^2}{n^2} + 3}\ri) - 3\ri]^{-1} \la{eq:b'}
\ea
Thus $S_{\ell}$ can be written down {\em analytically} using Eq. (\ref{eq:S_l}) (gives $S_l$ in terms of $\xi$), Eq. (\ref{eq:xi_l}) (gives $\xi$ in terms of $\b'$), Eq. (\ref{eq:b'}) (gives $\b'$ in terms of $K_{nn}$) and Eq. (\ref{eq:Knn}) (gives $K_{nn}$ in terms of $l/n$). After simplification, we have
\be
S_{\ell} = - \ln\le(1 - \f{\b'}{1+ \sq{1-\b'^2}}\ri) - \f{\b'}{1-\b' +\sq{1-\b'^2}} \ln\le(\f{\b'}{1 +\sq{1-\b'^2}}\ri) \la{eq:S_t}
\ee
where $\b'$ is given by Eq. (\ref{eq:b'}). Note that, no summation over $j$ is involved as there exists only one $\xi$ resulting from Eq. (\ref{eq:xi_l}) and Eq. (\ref{eq:b'}). It is easy to see that the resulting $S_{\ell}$ is essentially a function of $\ell/n$. Let us define a new variable $t = \ell/n$. 
The asymptotic expansion of $S_t$ near $t \rightarrow \infty$ is given by
\be
\lim_{t \rightarrow \infty} S_t = \f{0.0625 (3.7726 + 4 \ln t)}{t^4} - \f{(0.69315 + \ln t)}{t^6} + O(t^{-8}) \simeq \f{\ln t}{4\,t^4} \la{eq:asymS_t1}.
\ee
This matches exactly with \cite{1993-Srednicki-PRL} (see Eq. (\ref{eq:asymS_t2}) later).

\begin{figure}[!htb]
\begin{center}\hspace*{-24pt} 
\includegraphics[scale=0.5]{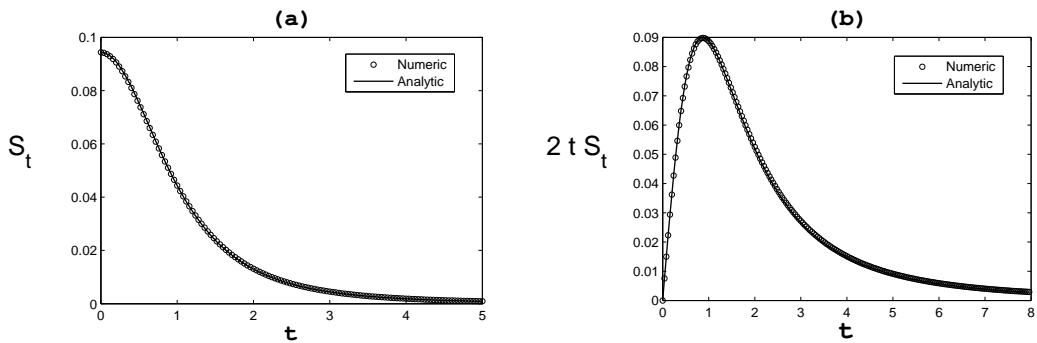}
\caption{(a) Plot of $S_t$ vs $t$ for $n = 50$. (b) Plot of $S_t$ vs $t$ for $n = 50$.}\label{fig:S_t}
\end{center}
\end{figure}
The functional dependence of $S_t$ on $t$, as given by Eq. (\ref{eq:S_t}), is shown by the continuous curve in Fig. \ref{fig:S_t}(a) (for $N=100$ and $n=50$). The small circles in Fig. \ref{fig:S_t} represent numerically computed $EE_{surf}$ without assuming $n,\ell\gg 1$. The agreement of analytic and numerical results seems satisfactory. Fig. \ref{fig:S_t}(b) shows the variation of the function `$2tS_t(t)$' which will be used later.


%
%
%

\begin{figure}[!htb]
\begin{center}\hspace*{-24pt} 
\subfigure [$~ n=20$]{
\includegraphics[scale=0.38]{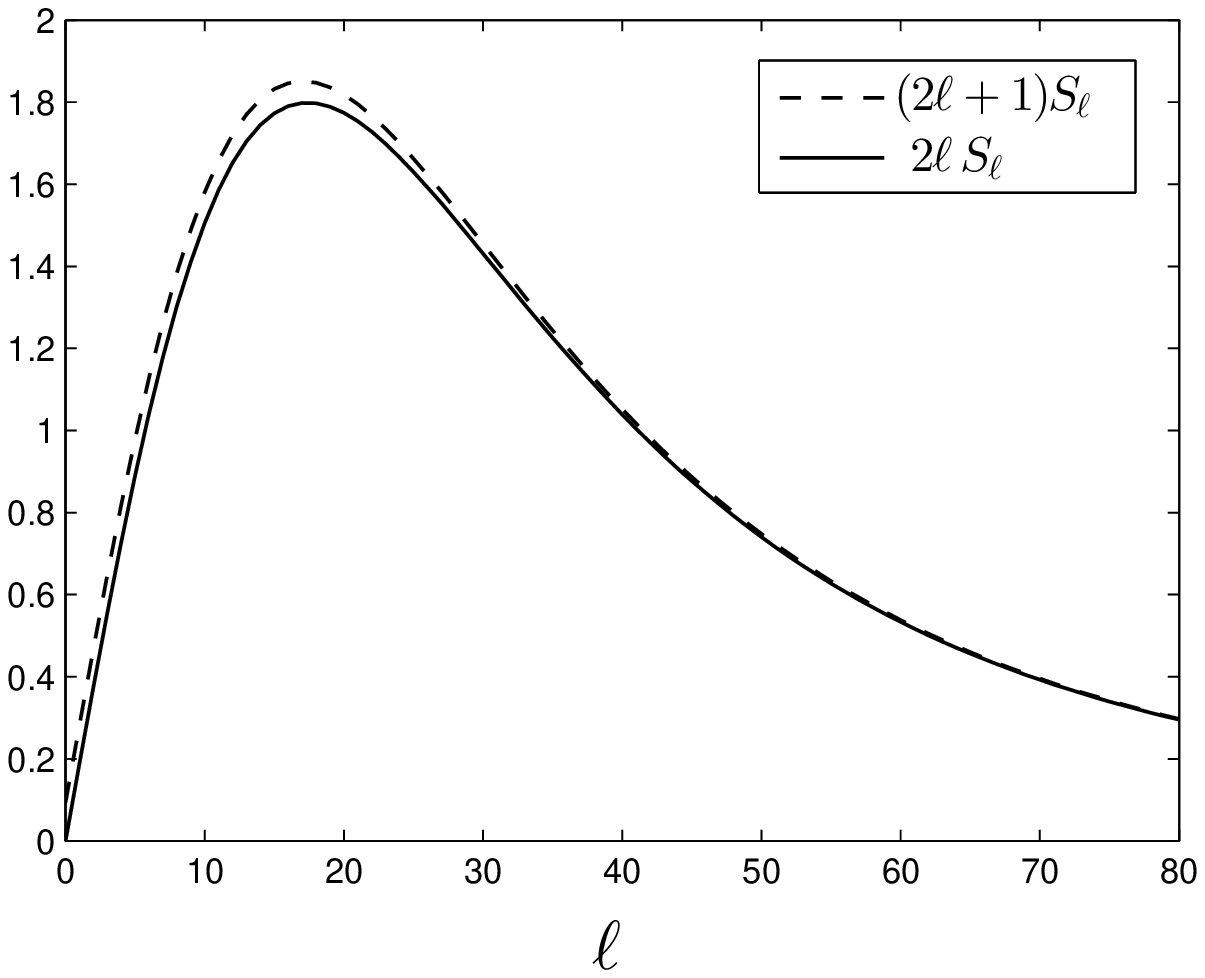} }
\subfigure [$~ n=50$]{
\includegraphics[scale=0.39]{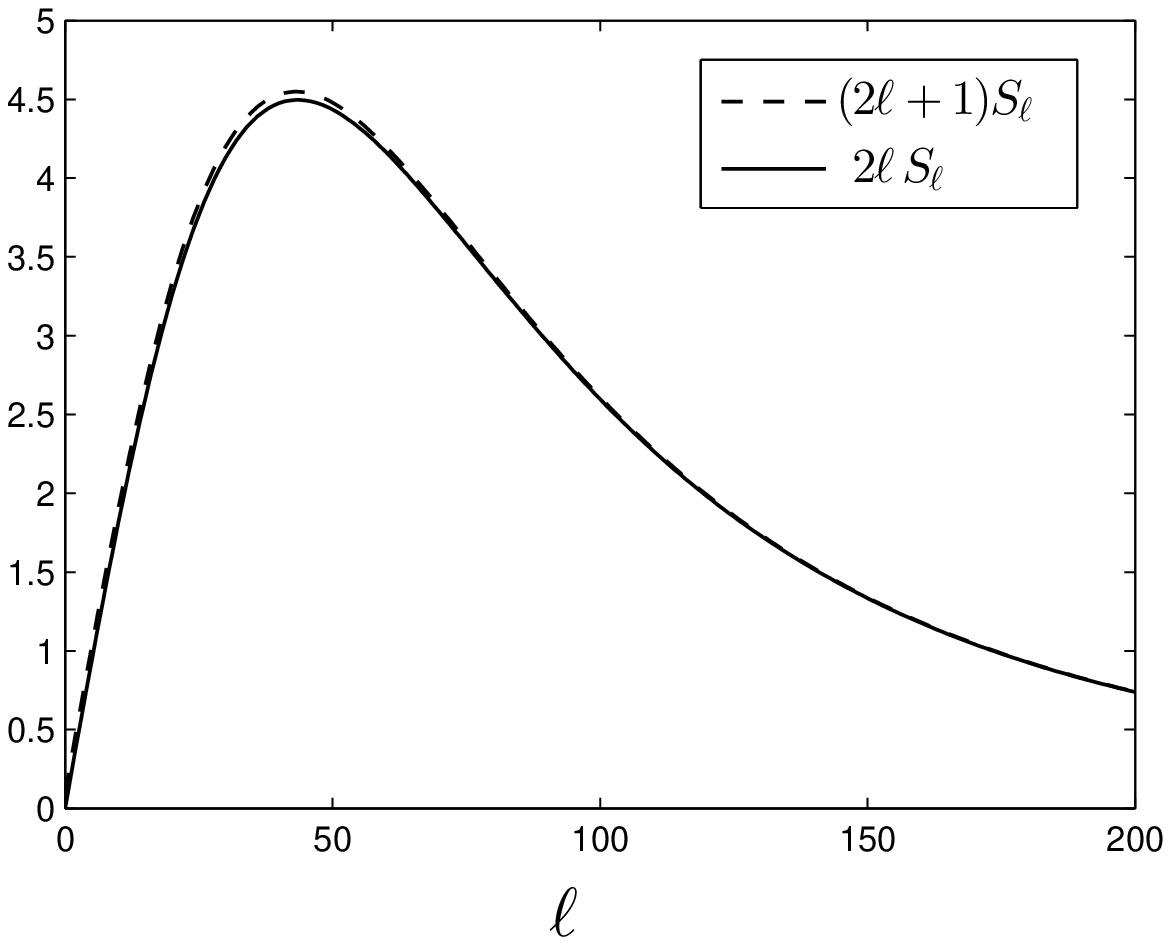} }
\subfigure [$~ n=80$]{
\includegraphics[scale=0.37]{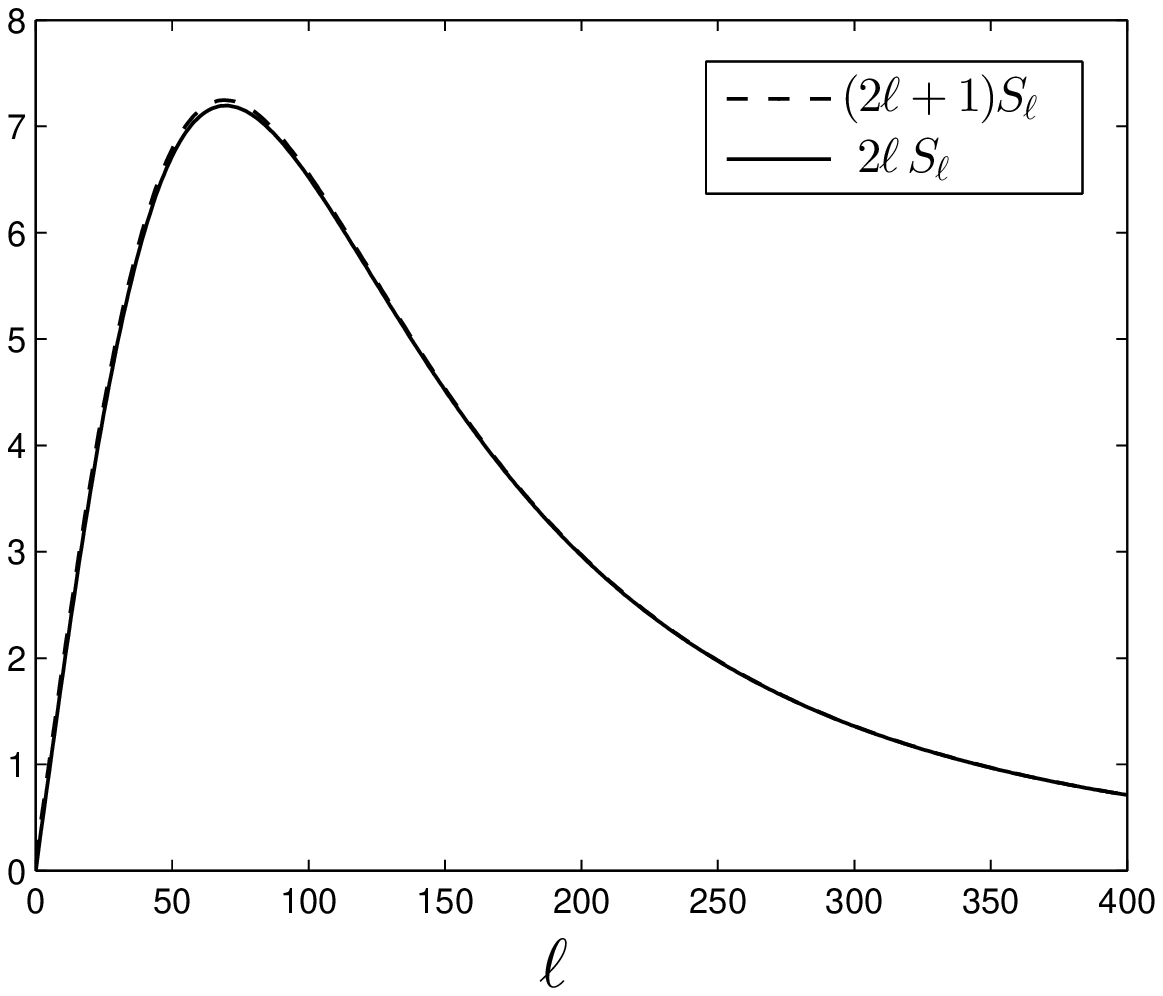} }
\caption{Comparison between `$(2\ell+1)\,S_{\ell}$' and `$2\ell\,S_{\ell}$' for $n = 20,50$ and $80$ with $N=100$.}\label{fig:compS_l}
\end{center}
\end{figure}

Now, we argue that for large $n$, Eq. (\ref{eq:S_tot}) approximates to
\be
EE_{surf} = \sum_{\ell=0}^{\infty}\, (2\ell + 1)S_{\ell}(\ell/n) \simeq \sum_{\ell=0}^{\infty}\, 2\ell\,S_{\ell}(\ell/n) \la{eq:S_tot_largel}
\ee
To visualise the amount of error resulted due to the above approximation we compare $(2\ell+1)S_{\ell}(\ell/n)$ and $2\ell\,S_{\ell}(\ell/n)$ vs $\ell$, where $S_{\ell}$ is given by Eq. (\ref{eq:S_t}), for $n = 20,50$ and $80$ (with $N=100$) in Fig \ref{fig:compS_l}.
Interestingly, Fig \ref{fig:compS_l} suggests two facts: (i) with increasing $n$, the relative error in the approximation made in Eq. (\ref{eq:S_tot_largel}) becomes increasingly negligible and (ii) $\ell \sim n$ modes contribute most to the total entropy which justifies the approximation made in Eq. (\ref{eq:S_tot_largel}) for large $n$ (which is indeed very large for a macroscopic system with $a \sim$ Planck length).

Eq. (\ref{eq:S_tot_largel}) further simplifies to
\ba
EE_{surf} &\simeq & \sum_{\ell=0}^{\infty}\, \triangle \ell\, 2\ell\,S_{\ell}(\ell/n);  ~~~~ \triangle \ell  = 1,\nonumber\\ 
&=& n^2 \underbrace{ \sum_{\ell=0}^{\infty}\, \triangle t\, 2t\,S_t(t)}_{c_2};  ~~~~ \triangle t = \f{\triangle \ell}{n}  = 1/n,  \nonumber\\
&=& c_2\, n^2  \equiv \f{c_2}{4\pi} \f{\mbox{horizon area}}{a^2}, \la{eq:S_surf}    
\ea
where $c_2$ is the pre-factor of the so-called `area' term. Eq. (\ref{eq:S_surf}) represents the well known area law and is the main result of this paper. 
For large $n$, one can further approximate the summation in Eq. (\ref{eq:S_tot_largel}) into an integration, leading to
\be
c_2 = 2 \int_0^\infty dt\,t \,S_{t}(t) = 0.2335.  \la{eq:c_2}    
\ee
The error in approximating the sum with the above integral is given by
\ba
\mbox{Error}(c_2) &\leq & \triangle t \le(|[2t \,S_{t}]_{t=0} - [2t \,S_{t}]_{t=t_0}| + |[2t \,S_{t}]_{t=t_0} - [2t \,S_{t}]_{t \rightarrow \infty}|\ri) \\
&=& \f{1}{n} [4t \,S_{t}]_{t=t_0}
\ea
where $t_0$ denotes location of the maximum for the function `$t\,S_t$' (see Fig. \ref{fig:S_t}) and we have used the fact that $S_t \rightarrow 0$ as $t \rightarrow \infty$. Using Eq. (\ref{eq:S_t}), we find that $t_0 \sim 0.87$ and max$(tS_t) \sim 0.045$. Thus we get
\be
\mbox{Error}(c_2) \leq \f{0.18}{n}.
\ee
Further, with increasing $n$, $\mbox{Error}(c_2) \rightarrow 0$ (e.g. for $n=50$, $\mbox{Error}(c_2) \leq 0.0036$).




Let us compare $EE_{surf}$ with $EE_{tot}$ where the DoF away from the horizon are also contributing however small. Consider a $s\times s$ window instead of a $2\times2$ window in Eq. (\ref{eq:Kij_simp}) with $2\leq s<<n$. Now, one can determine $EE_{surf}$ numerically as a function of $s$. As already reported in [14], $EE_{surf}(s)$ contributes 85\%, 94\%, 97\% and 98\% of $EE_{tot}$ with $s = 4,6,8,10$ respectively. Thus $EE_{surf}(s) \sim EE_{tot}$ even with $s<<n$. Note that,  for  $\forall s<<n$ we can again write $K_{i,j} \forall i,j$ as functions of $\ell/n$ which implies that $S_{\ell}$ is again a function of $\ell/n$ only. Thus the functional form of $EE_{surf}(s)$ is once again given by Eq.(20). However, the coefficient $c_2$ can not be derived in an integral form anymore as in Eq. (20) using this algorithm. 
 

\subsection{$EE_{surf}$ in any dimension}
It is straightforward to extend these analysis to any dimension where the {\it horizon} is spherically symmetric. In \cite{Braunstein:2011sx}, authors have computed numerically $EE_{tot}$ (defined to be the R{\'e}nyi entropy) for D dimensional spherical entangling surfaces. 
In such scenario, the `radial part' of the Hamiltonian is given by
\be
H_{\ell,m_i} = \int_0^\infty dr \le[\pi_{\ell,m_i}^2 + r^D \le\{\pa_r \le(\f{\phi_{\ell,m_i}}{r^{D/2}}\ri)\ri\}^2 + \f{\ell(\ell + D-1)}{r^2} \phi_{\ell,m_i}^2\ri] \la{eq:Ham-gen}
\ee
where $m_i$ represents angular momentum for $i$-th azimuthal angular coordinate where $i+1=D$.
After discretization one gets the matrix elements of $K_{ij}$ as,
\be
K_{jj} = j^{-D} \le[\le(j+\f{1}{2}\ri)^D\ + \le(j+\f{1}{2}\ri)^D + j^{D-2}\ell(\ell +D-1)\ri], ~~ K_{j,j+1} = \f{\le(j+\f{1}{2}\ri)^D}{j^D j^{D+1}} \la{eq:Kij-gen}
\ee
The resulting asymptotic expression for $\xi_l$ for $\ell \gg N$ is given by \cite{Braunstein:2011sx},
\be
\xi_\ell \simeq \f{1}{2^{2D+2}}\f{(2n+1)^{2D-2}\{n(n+1)\}^{3-D}}{\ell^2(\ell +D-1)^2}. \la{eq:xi_l-gen}
\ee
Thus for $\ell\gg D$ and $n\gg 1$ we have
\be
\xi_\ell \equiv \xi_t \simeq \f{1}{16t^4}
\ee
which is, notably, independent of $D$ and matches with the corresponding result in \cite{1993-Srednicki-PRL}. Thus the asymptotic behaviour of $S_\ell$ or $S_t$ is same in any dimension and given by
\ba
\lim_{t\rightarrow \infty} S_t &\simeq & \lim_{t\rightarrow \infty} \xi_t (1 - \ln \xi_t)\\
&= & \lim_{t\rightarrow \infty} \f{1 + \ln (16t^4)}{16t^4}\\
&\simeq & \lim_{t\rightarrow \infty} \f{\ln t}{4\,t^4} \rightarrow 0 \la{eq:asymS_t2}
\ea
which is consistent with Eq. (\ref{eq:asymS_t1}) and Fig. \ref{fig:S_t}.
Let us now derive $EE_{surf}$ for any `$D$'.
Again, for $j = n \gg 1$ and $l>>D$, we get back Eq. (\ref{eq:Knn}) which is independent of $D$. This implies that the resulting $S_t$ for $EE_{surf}$ in {\em any} dimensions is again given by Eq. (\ref{eq:S_t}).

Thus if the horizon is a spherical hypersurface of dimension $D$, Eq. (\ref{eq:S_tot}) is replaced by,
\be
EE_{surf}^{(D)} = \sum_{\ell = 0}^\infty g_{{}_\ell}^{(D)} \,S_t; \hspace{1cm} \mbox{where}~~~~ g_{{}_\ell}^{(D)} = \f{(2\ell +D-1)(\ell +D-2)!}{(D-1)! \, \ell!},
\ee
where $g_{{}_\ell}^{(D)}$ is the degeneracy factor.
Again for $\ell \gg D$, $g_\ell \sim 2\ell^{D-1}/(D-1)!$. Thus we have
\ba
EE_{surf}^{(D)} &\simeq & \sum_{\ell = 0}^\infty \triangle \ell\, \f{2}{(D-1)!} \ell^{D-1}\, S_t; ~~~~~\triangle \ell  = 1\\
&\simeq & n^{D} \underbrace{ \sum_{\ell = 0}^\infty \triangle t\,\f{2}{(D-1)!}t^{D-1} \, S_t}_{c_{{}_D}}; ~~~~~\triangle t  = \f{1}{n} \\
&\simeq & c_{{}_D} n^{D}. \la{eq:S_D}
\ea
This is the area law in higher dimensions that was first demonstrated in \cite{Braunstein:2011sx}.
Let us again approximate the summation in the definition of $c_{{}_D}$ with an integration for large $n$, such that
\be
c_{{}_D} \sim \int_0^\infty dt\,\f{2}{(D-1)!}t^{D-1} \, S_t. \la{eq:c_D}
\ee
Using Eq. (\ref{eq:asymS_t1}), we see that, as $t \rightarrow \infty$, the intgrand in Eq. (\ref{eq:c_D}) goes as $t^{D-5}\ln t$. This implies that $c_{{}_D}$ does not converge to a finite value for $D\geq 4$ \cite{1993-Srednicki-PRL,Braunstein:2011sx}. To get a finite result one can either use the definition of R{\'e}nyi entropy \cite{Braunstein:2011sx}, where one can adjust a free parameter such that the above integration converges or one can use an angular momentum cut-off for $\ell_{max}$ and $m_{max}$ that is consistent with the linear momentum cut-off `$a^{-1}\pi$', i.e.
\be
\ell_{max}, m_{max} \sim r \times a^{-1}\pi \sim n\, a \times a^{-1}\pi \sim \pi \, n.
\ee

%

%

\begin{figure}[!htb]
\begin{center}\hspace*{-24pt} 
\includegraphics[scale=0.5]{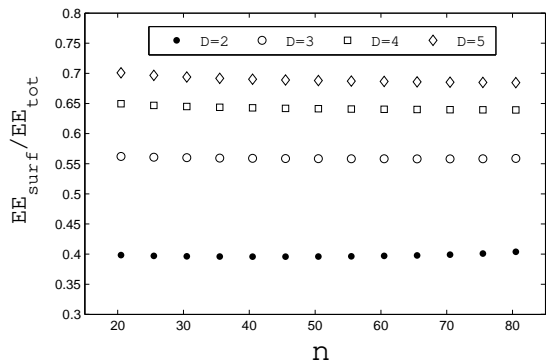}
\caption{$EE_{surf}/EE_{tot}$ vs $n$ in different $D+1$ spacetime dimensions.}\label{fig:01}
\end{center}
\end{figure}

One apparently simple fact can be learned from the variation of the ratio of $EE_{surf}$ and $EE_{tot}$ (computed numerically using MATLAB for $N=100$ and $\ell_{max}=n$ as an universal cut-off in any $D$) in different spacetime dimensions shown in Fig. \ref{fig:01}. Note that, as $D$ increases this ratio also increases and tends to unity i.e. $EE_{surf}$ saturates $EE_{tot}$ with increasing dimension. This is because the ratio of the surface area of a $(D-1)-surface$ and volume of a $D-sphere$ of radius $\cal R$ is given by $D/\cal R$. As the relative number of surface DoF on the horizon with respect to the volume DoF increases with increasing spatial dimension and the resulting $EE_{surf}/EE_{tot}$ also increases with increasing $D$. This further implies that Eq. (\ref{eq:S_D}) becomes a more and more exact formula for EE in higher dimensions.

\subsection{Numerical analysis}
\begin{figure}[!htb]
\begin{center}\hspace*{-24pt} 
\includegraphics[scale=0.5]{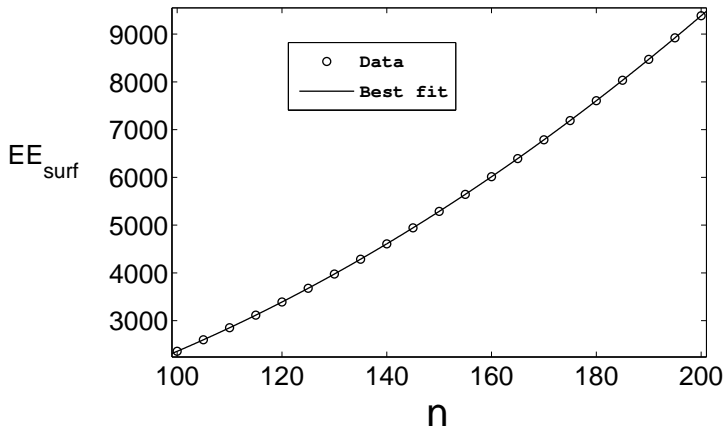}
\caption{(a) Fitting of numerically computed $EE_{surf}$ by Eq. (\ref{eq:Theisen-fit}).} \label{fig:Esurf_fit}
\end{center}
\end{figure}

In the geometric approaches, a log dependent term appears even in flat spacetime when the entangling surface has an extrinsic curvature \cite{Solodukhin}. Thus, anticipating the presence of a logarithmic correction, in \cite{Theisen-2009}, authors have numerically fitted $EE_{tot}$ with the following function
\be
F = c_2 (n+ 0.5)^2 + c_{log} \log(n+0.5) + c \la{eq:Theisen-fit}
\ee
with best fit values as 
\be
c_2 = 0.2954; ~~ c_{log} = - 0.011;~~ c = 0.035. \la{eq:bestfit-Theisen}
\ee
Thus $c_{log}$ was found to be consistent with the predicted value for an extremely charged black hole from the geometric approaches \cite{Solodukhin} that is `$-1/90$' \footnote{The corresponding value for a Schwarzschild black hole is `$1/45$' \cite{Solodukhin}.}. 
Accordingly, in Fig. \ref{fig:Esurf_fit}, we fit numerically computed (using MATLAB) $EE_{surf}$ with Eq. (\ref{eq:Theisen-fit}). We have used $N=300$ and $n = 100 - 200$. To decrease the computational error even further, we have set the cut-off $\ell_{max}$ at a percentage error of $10^{-7}\%$, i.e. the relative error is $[EE_{surf}(\ell_{max}) - EE_{surf}(\ell_{max}-1)]/EE_{surf}(\ell_{max})  <  10^{-9}$.

The resulting best fit values for $EE_{surf}$ are given by
\be
c_2 = 0.2334 \pm 10^{-6}; ~~ c_{log} = 0.176 \pm 0.035;~~ c = -0.75 \pm 0.15. \la{eq:bestfit-Esurf}
\ee
Best fit $c_2$ matches with our predicted value in Eq. (\ref{eq:c_2}) and thus confirms the reliability of the numerical results. There is a `positive' logarithmic term present too. This is simply the result of only considering interaction among near horizon DoF. 
Comparison between Eq. (\ref{eq:bestfit-Esurf}) and Eq. (\ref{eq:bestfit-Theisen}) implies that 
the role of DoF away from the horizon is to increase the value of $c_2$ and also to compensate for the negative value of $c_{log}$ in Eq. (\ref{eq:bestfit-Theisen}).

\section{Discussions}
Our aim was to {\em analytically} compute the dominant terms in EE using reasonable approximations in the non-geometric formalism developed in \cite{1986-Bombelli.etal-PRD,1993-Srednicki-PRL}. We have defined $EE_{surf}$ to be the entropy due to entanglement among near horizon DoF which captures the leading contributions. In the following, we summarise the key results.

\begin{itemize}

\item Analytic derivation of $EE_{surf}$, given by Eq. (\ref{eq:S_surf}), is the main result of this article. This area law was expected as we have considered the contribution of the near horizon DoF only. However, the technical reason behind emergence of the area law is that the entropy contributed by the individual modes depend on the ratio $\ell/n$.

\item $\ell \sim n$ modes contributes most to the total entropy.

\item A logarithmic sub-leading term, with a positive coefficient is found numerically. This implies that the DoF away from the horizons contribute to the total entropy in such a way that the final coefficient of the logarithmic term becomes negative and the proportionality constant with the area term increases.


\item The area law is shown to hold in higher dimensions too. However the coefficient diverges when the dimension of the entangling surface is $\geq 4$. This problem can be solved, if one uses the same lattice cut-off for angular directions as is used for radial direction, one gets $\ell_{max} \sim \pi n$ which results in a finite entropy. 

\item  $EE_{surf}$ saturates $EE_{tot}$ with increasing dimensions.

\end{itemize}

In the so-called non-geometric approach \cite{1986-Bombelli.etal-PRD,1993-Srednicki-PRL}, computation of EE has been mostly done numerically in various scenarios to extend the known area law to the more generic cases \cite{Das:2008sy}. Our work provides an analytic recipe to look through the numerical complexities and is proved to be useful to extend the area law to higher dimensions too.
It will be worthwhile to attempt computing $EE_{surf}$ for non-trivial horizon geometries \cite{Kumar:2014kua}, which are relevant in the context of `black' objects in higher dimensions such as black rings \cite{Emparan:2001wn,Emparan:2006mm}.
We hope to report on these issues elsewhere.

%
%
%
%

\section*{Acknowledgments}

The author thanks K. Banerjee, S. Braunstein, S. Das,  A. Lahiri, P. Sarkar, S. Shankaranarayanan, S. Theisen and R. Tibrewala for useful comments and discussions.

%


\end{document}